# Creating high-dimensional topological physics using a single ring resonator

Dali Cheng,[1,2] Heming Wang,[1] Charles Roques-Carmes,[1] Janet Zhong,[1,3] and Shanhui Fan[1,2,3*]

[1]*Ginzton Laboratory, Stanford University, Stanford, California 94305, USA*
[2]*Department of Electrical Engineering, Stanford University, Stanford, California 94305, USA*
[3]*Department of Applied Physics, Stanford University, Stanford, California 94305, USA*
[*]shanhui@stanford.edu

**Abstract**

In spaces of three or more dimensions, there exists topological physics of significant richness that has no lower-dimensional counterparts. To experimentally explore high-dimensional physics, it is advantageous to augment the physical space with synthetic dimensions. An emerging approach is to use a *single* modulated photonic ring resonator to form *multiple* synthetic frequency dimensions. However, all the high-dimensional Hamiltonians experimentally demonstrated in this approach are topologically trivial. Here we propose a general scheme to create high-dimensional topological physics using multiple synthetic frequency dimensions in a single ring resonator. This scheme utilizes specifically designed mode-selective modulations and mode conversions to create non-trivial topology. As examples, we numerically demonstrate a three-dimensional, two-band model exhibiting Weyl points and topological insulator phases, and a five-dimensional, four-band model exhibiting Yang monopoles and Weyl surfaces. Our results will facilitate the experimental studies and future applications of topological physics in high dimensions.

**Introduction**

In topological physics, the exploration of high-dimensional models is of theoretical interest since these models exhibit physics that cannot be found in lower-dimensional systems. A well-known example is the Weyl point, a monopole of the U(1) Berry curvature that only exists in systems of at least three dimensions [1]. As another example, five-dimensional systems can exhibit Yang monopoles and interlinked Weyl surfaces that do not exist in lower-dimensional spaces [2,3].

The importance of these theoretical results of high-dimensional topology motivates their experimental realization. To construct experimental systems beyond three dimensions, it is necessary to utilize the concept of synthetic dimensions, which supplements the spatial dimensions with additional degrees of freedom to form higher-dimensional spaces. Even in three-dimensional systems, the use of synthetic dimensions is also beneficial to reduce the experimental complexity. Systems with synthetic dimensions have been extensively studied in the past decades in atomic [4–13], photonic [14–25], and acoustic systems [26]. As an important example showing the explorations of high-dimensional systems, the four-dimensional quantum Hall effect has been studied in synthetic dimensions [10,11,27–29].

A prominent approach to create synthetic dimensions in photonics is to form synthetic frequency dimensions by coupling modes at different frequencies in modulated ring resonators [27,30–50]. There have been theoretical proposals aiming to explore topological physics in or beyond three dimensions using synthetic frequency dimensions, including Weyl points in three-dimensional systems [33,51], three-dimensional topological insulators [52], the four-dimensional quantum Hall effect [27], and high-order topological insulators in high dimensions [53]. All these proposals require a ($D$−1)-dimensional array of ring resonators to implement a $D$-dimensional lattice, and therefore $D \leq 4$, and the use of multiple rings results in significant implementation complexity. Alternatively, it has been theoretically noted that high-dimensional lattices can be formed in a single modulated ring resonator using multiple modulation frequencies [54]. However, all the Hamiltonians in three or more dimensions implemented using this approach are topologically trivial [40,45].

In this paper, we propose a general scheme to create topological physics in arbitrarily high dimensions, using synthetic frequency dimensions in a single ring resonator formed by a multi-mode waveguide. The high-dimensional lattice is generated by high-order modulations in the ring resonator, and multiple bands are created using different spatial modes in the waveguide. We show that a general form of multi-dimensional, multi-band Hamiltonian can be implemented by designing the mode-selective modulations and mode conversions in the ring resonator. As illustrative examples, we implement Weyl points and topological insulator phases in three dimensions, and Yang monopoles and Weyl surfaces in five dimensions. The band structures of these systems are measurable through the time-dependent spectra of the ring resonator. Our proposed scheme provides a versatile and programmable platform to study high-dimensional topological physics.

**Results**

We aim to implement a $D$-dimensional, $S$-band Hermitian lattice model that takes the following form in the reciprocal space:

$$H(\mathbf{k}) = \sum_{g=1}^{G} f_g(\mathbf{k}) \gamma_g$$

(1)

where $\mathbf{k} = [k_1, \ldots, k_D]^\mathrm{T}$ is the momentum vector in the $D$-dimensional space, and $f_g(\mathbf{k})$ is a real function that has period $2\pi$ in each of the dimensions of $\mathbf{k}$. We keep $f_g(\mathbf{k})$ as a general function here and will later provide its detailed form when discussing specific examples. $\{\gamma_1, \ldots, \gamma_G\}$, with $G = S^2 - 1$, are generators of the SU($S$) group and form a complete basis of the $S \times S$ traceless Hermitian matrices. One standard choice of $\{\gamma_1, \ldots, \gamma_G\}$ consists of generalized Gell–Mann matrices [55]. We assume that the Hamiltonian Eq. (1) describes a $D$-dimensional cubic lattice with uniform onsite potentials and uniform nearest-neighbor tight-binding couplings. The corresponding real-space Hamiltonian can be found in the Supplemental Materials. Eq. (1) is widely used for describing theoretical models in the studies of high-dimensional topological physics [1–3,56–58].

Fig. 1 shows the setup to implement Eq. (1) in a single ring resonator. We consider a ring resonator formed by a multi-mode waveguide that supports a total of $S$ degenerate spatial modes. These spatial modes act as the pseudo-spin degree of freedom in the multi-band model. The multi-mode waveguide is split into $G$ different branches in parallel which then combine back. Each branch consists of a pair of mode converters sandwiching a mode-selective modulation section [Fig. 1(b)]. In the mode-selective modulation section, each spatial mode is split into a sub-branch consisting of a single-mode waveguide. Within each sub-branch there is an electro-optic phase modulator with time-dependent transmittance $\tau_{gs}(t)$. Here the subscript $g$ labels the branch and $s$ labels the spatial mode. The ring resonator supports a set of resonant frequencies forming a lattice with the free spectral range $\Omega_R$. We assume that all spatial modes have the same free spectral range when the modulations and mode conversions are absent.

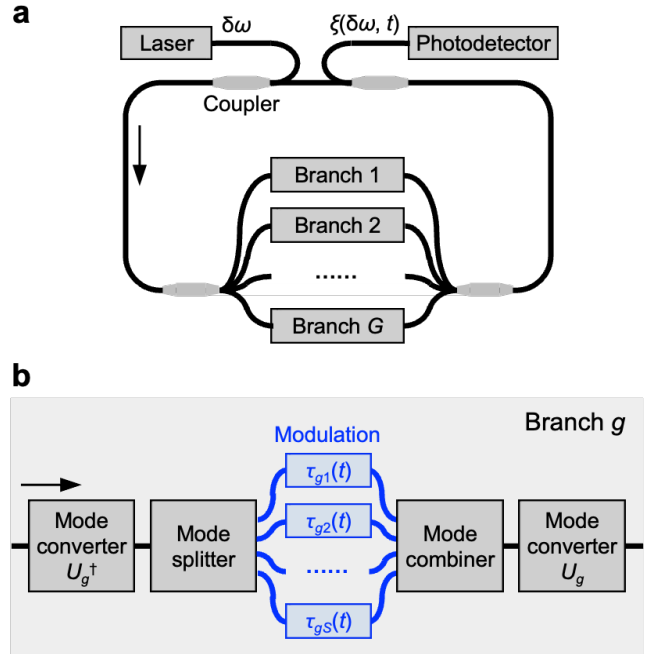

Fig. 1. Setup to implement a multi-dimensional, multi-band lattice Hamiltonian. (a) Schematic of the ring resonator including $G$ different branches. (b) Detailed structure of the $g$-th branch, including a pair of mode converters and $S$ mode-selective modulators. Black lines represent multi-mode waveguides, and blue lines represent single-mode waveguides.

To implement the Hamiltonian Eq. (1) using the setup in Fig. 1, we choose the transmittances $\tau_{gs}(t)$ and the unitary transformations $U_g$ as follows: Each term in Eq. (1) is implemented by a branch in Fig. 1. For the $g$-th term $f_g(\mathbf{k})\gamma_g$, we first diagonalize $\gamma_g = U_g \Lambda_g U_g^\dagger$. The diagonal matrix $\Lambda_g$ contains the eigenvalues $\{\lambda_{g1}, \ldots, \lambda_{gS}\}$ that are arranged in the order from mode 1 to mode $S$ in the waveguide. Here $U_g$ is unitary because $\gamma_g$ is Hermitian, and is implemented by the mode converters. The transmittances of the modulators are then set as $\tau_{gs}(t) = \exp[i\kappa T_R \lambda_{gs} f_g(\mathbf{k}(t))]$, where $\kappa$ is the modulation strength, $T_R = 2\pi/\Omega_R$ is the roundtrip time of the resonator, and $\mathbf{k}(t) = [k_1(t), \ldots, k_D(t)]^T$ with the $d$-th element being $k_d(t) = M^{d-1}\Omega_R t + \varphi_d$. Here $M$ is an integer, and we take $M = 3$ for all the numerical examples in the paper. The phase variable $\varphi_d$ is introduced to facilitate the probing of the band structure of the Hamiltonian. Note that the modulation signal $\tau_{gs}(t)$ is periodic and contains frequency components $M^{d-1}\Omega_R$ ($1 \leq d \leq D$). With the modulation frequency $M^{d-1}\Omega_R$, the $m$-th resonance is coupled to the ($m \pm M^{d-1}$)-th resonances. Hence, we create a one-dimensional tight-binding lattice along the frequency axis with long-range couplings. Such a one-dimensional lattice can be mapped to a $D$-dimensional lattice with twisted boundary conditions [54], which has a finite width of $M$ sites along the first ($D - 1$)-dimensions, and is infinite along the $D$-th dimension. An illustration of this mapping can be found in the Supplemental Materials. Finally, in the weak-modulation limit $\kappa T_R \ll 1$, the Hamiltonian implemented by all the $G$ branches is the average of those by individual branches $f_g(\mathbf{k})\gamma_g$. This is because the terms representing the synergistic effects of multiple branches are of higher orders of $\kappa T_R$ and can thus be neglected. A detailed derivation can be found in the Supplemental Materials showing that the evolution dynamics of the light field in the resonator is governed by a Schrödinger-like equation described by the Hamiltonian Eq. (1), up to a constant factor.

Related to our scheme above, there have been proposals to realize multi-dimensional [54,59] and multi-band [60,61] lattice models, using synthetic frequency dimensions in a single or a few ring resonators. However, the Hamiltonian in the general form of Eq. (1) has not been previously implemented, and the physics of high-dimensional topological semimetals and topological insulators has not been demonstrated in the synthetic frequency dimension in a single ring resonator.

We use the methods discussed in [59,62] to probe the band structure of the Hamiltonian Eq. (1). The resonator is excited by a continuous-wave laser, with frequency detuning $\delta\omega$ from one of the resonant frequencies. The steady-state light intensity $\xi$ inside the resonator is sampled and measured by a photodetector, as a function of $\delta\omega$ and time $t$. We refer to $\xi(\delta\omega, t)$ as the time-dependent spectrum of the resonator. In a synthetic frequency lattice, the time $t$ is viewed as the momentum $\mathbf{k}$, and the peaks in $\xi(\delta\omega, t)$ correspond to the band energies $E(\mathbf{k})$ of the Hamiltonian sampled along a one-dimensional line in the $D$-dimensional Brillouin zone: $k_d = M^{d-1}k_1 + \varphi_d$ (mod $2\pi$), $2 \leq d \leq D$, $\varphi_1 = 0$. These $\varphi_d$'s are phases of the frequency component $M^{d-1}\Omega_R$ in the modulation signal $\tau_{gs}(t)$, and by tuning them one can fully measure the band structure across the $D$-dimensional Brillouin zone.

In the following, we implement as examples a three-dimensional Hamiltonian $H_{3D}$ exhibiting either a three-dimensional topological semimetal phase with Weyl points or a three-dimensional topological insulator phase, as well as a five-dimensional Hamiltonian $H_{5D}$ exhibiting Yang monopoles and Weyl surfaces. We also perform simulations on their band structure measurements by calculating the time-dependent spectra $\xi(\delta\omega, t)$ that highlight the degeneracies and band gaps.

For the studies of three-dimensional topological semimetals or topological insulators, we consider

$$H_{3D} = \sin(k_1)\,\sigma_1 + \sin(k_2)\,\sigma_2 + (2 + u - \cos k_1 - \cos k_2 - \cos k_3)\sigma_3$$

(2)

where $\sigma_{1(2,3)}$ are the 2×2 Pauli matrices, and $u$, a real parameter, is the onsite potential. Written in the form of Eq. (1), we have $f_1(\mathbf{k}) = \sin k_1$, $f_2(\mathbf{k}) = \sin k_2$, and $f_3(\mathbf{k}) = 2 + u - \cos k_1 - \cos k_2 - \cos k_3$. When $-1 < u < 1$, $H_{3D}$ supports a pair of Weyl points of opposite charges at $k_1 = k_2 = 0$, $k_3 = \pm\cos^{-1}(u)$. Figs. 2(a) and 2(b) show the case of $u = 0$. Fig. 2(a) shows the locations of the Weyl points. Fig. 2(b) displays the band structure at $k_2 = 0$, and the linear dispersions are evident in the vicinity of $k_3 = \pm\pi/2$. When the physical system described by this Hamiltonian is truncated, Fermi arc surface states appear with their dispersions connecting the projections of the two Weyl points onto the surface. The Hamiltonian $H_{3D}$ describes a topological insulator when $u < -1$, and a trivial insulator when $u > 1$.

In the five-dimensional reciprocal space, there are two types of generalizations of Weyl points: the Yang monopole and the Weyl surface [2,3]. A Yang monopole is a zero-dimensional monopole of the SU(2) Berry curvature, and a Weyl surface is a two-dimensional manifold as the source of the U(1) Berry curvature. Both types of degeneracies are characterized by second Chern numbers, and can be found in the following Hamiltonian:

$$H_{5D} = \sin(k_1)\,\gamma_1 + \sin(k_2)\,\gamma_2 + \sin(k_3)\,\gamma_3 + \sin(k_4)\,\gamma_4$$
$$+ (4 + u - \cos k_1 - \cos k_2 - \cos k_3 - \cos k_4 - \cos k_5)\gamma_5 + b\gamma_6$$

(3)

where $u$ and $b$ are real numbers describing two types of onsite potentials. The 4×4 gamma matrices are constructed by the direct product of Pauli matrices: $\gamma_1 = \sigma_3 \otimes \sigma_1$, $\gamma_2 = \sigma_3 \otimes \sigma_2$, $\gamma_3 = \sigma_3 \otimes \sigma_3$, $\gamma_4 = \sigma_1 \otimes \sigma_0$, $\gamma_5 = \sigma_2 \otimes \sigma_0$, $\gamma_6 = i[\gamma_3, \gamma_4]/2$. Written in the form of Eq. (1), $f_g(\mathbf{k}) = \sin k_g$, $1 \leq g \leq 4$, $f_5(\mathbf{k}) = 4 + u - \cos k_1 - \cos k_2 - \cos k_3 - \cos k_4 - \cos k_5$, and $f_6(\mathbf{k}) = b$. When $u = b = 0$, $H_{5D}$ supports a pair of Yang monopoles at $k_1 = k_2 = k_3 = k_4 = 0$, $k_5 = \pm\pi/2$, with a second Chern number of $\pm 1$ [Fig. 2(c)]. When $0 < b < 1$ and $u = 0$, the system exhibits three Weyl surfaces:

Weyl surfaces 1, 3:     $k_3 = k_4 = 0$, $\sin^2 k_1 + \sin^2 k_2 + (2 - \cos k_1 - \cos k_2 - \cos k_5)^2 = b^2$

Weyl surface 2:     $k_1 = k_2 = 0$, $\cos k_3 + \cos k_4 + \cos k_5 = 2$

(4)

These surfaces are linked to each other, as shown in Fig. 2(d) on the plane $k_2 = k_4 = 0$ when $b = 0.5$. Both the Yang monopoles and the Weyl surfaces are associated with topologically protected surface states when the lattice is truncated [63].

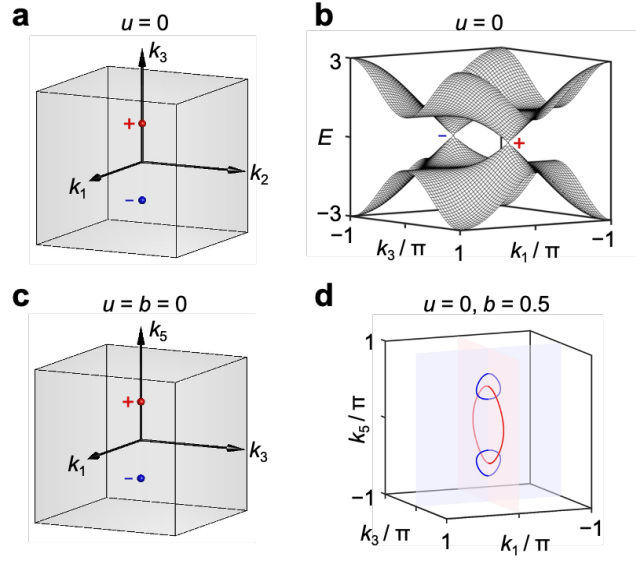

Fig. 2. Properties of the lattice models $H_{3D}$ and $H_{5D}$. (a) Locations of the Weyl points of $H_{3D}$ in the three-dimensional Brillouin zone when $u = 0$. (b) Band structure of $H_{3D}$ plotted as a function of $k_1$ and $k_3$, with $u = 0$ and $k_2 = 0$. (c) Locations of the Yang monopoles of $H_{5D}$ when $u = b = 0$. (d) Geometries of the Weyl surfaces of $H_{5D}$ when $u = 0$ and $b = 0.5$. Weyl surfaces are plot in the three-dimensional space of $(k_1, k_3, k_5)$ with $k_2 = k_4 = 0$, where these Weyl surfaces appear as lines, to show their linking behaviors. Weyl surfaces 1, 3 (2) are shown in blue (red), and the semi-transparent blue (red) planes at $k_3 = 0$ ($k_1 = 0$) are added as references.

To implement $H_{3D}$ with our proposed scheme in Fig. 1, we choose $D = 3$, $S = 2$, $G = 3$, and the unitary transformations $U_g$ and mode-selective modulations $\tau_{gs}(t)$ as in Table 1. Fig. 3 shows the band structures of the Hamiltonian $H_{3D}$. In Fig. 3(a), we show the theoretical band structures along the one-dimensional sampling line: $k_2 = 3k_1 + \varphi_2$ (mod $2\pi$), $k_3 = 9k_1 + \varphi_3$ (mod $2\pi$), by diagonalizing $H_{3D}$. We keep $\varphi_2 = 0$ in Fig. 3. In the first two panels $u = -1.5$, $H_{3D}$ describes a three-dimensional topological insulator, and therefore the band structure is gapped regardless of the choice of $\varphi_3$. In the third panel $u = 0$, and the one-dimensional sampling line touches the Weyl point at $k_1 = k_2 = 0$, $k_3 = \pi/2$ when $\varphi_3 = \pi/2$. This Weyl point corresponds to the band crossing at $k_1 = 0$ in the third panel of Fig. 3(a). In the last panel, $u = 0.5$. The Weyl point moves to $k_1 = k_2 = 0$, $k_3 = \pi/3$, and the band crossing is evident when $\varphi_3 = \pi/3$. These theoretical band structures can be resolved by simulating the time-dependent spectrum $\xi(\delta\omega, t)$ of the resonator. In Fig. 3(b), we calculate and show $\xi(\delta\omega, t)$ using the mode conversions and modulations in Table 1; see Supplemental Materials for methods of the calculation. The resonant features in $\xi(\delta\omega, t)$ closely match the theoretical band structures in Fig. 3(a), exhibiting band crossings or band gaps for different combinations of parameters.

Table 1. Modulations and mode conversions to implement $H_{3D}$. $\sigma_0$ is the 2×2 identity matrix.

| | $\tau_{g,s=1}(t)$ | $\tau_{g,s=2}(t)$ | $U_g$ |
|---|---|---|---|
| $g = 1$ | $\exp[+i\kappa T_R \sin(\Omega_R t)]$ | $\exp[-i\kappa T_R \sin(\Omega_R t)]$ | $\frac{1}{\sqrt{2}}\begin{bmatrix} 1 & -1 \\ 1 & 1 \end{bmatrix}$ |
| $g = 2$ | $\exp[+i\kappa T_R \sin(M\Omega_R t + \varphi_2)]$ | $\exp[-i\kappa T_R \sin(M\Omega_R t + \varphi_2)]$ | $\frac{1}{\sqrt{2}}\begin{bmatrix} -i & -i \\ 1 & -1 \end{bmatrix}$ |
| $g = 3$ | $\exp[+i\kappa T_R (2 + u - \cos(\Omega_R t) - \cos(M\Omega_R t + \varphi_2) - \cos(M^2\Omega_R t + \varphi_3))]$ | $\exp[-i\kappa T_R (2 + u - \cos(\Omega_R t) - \cos(M\Omega_R t + \varphi_2) - \cos(M^2\Omega_R t + \varphi_3))]$ | $\sigma_0$ |

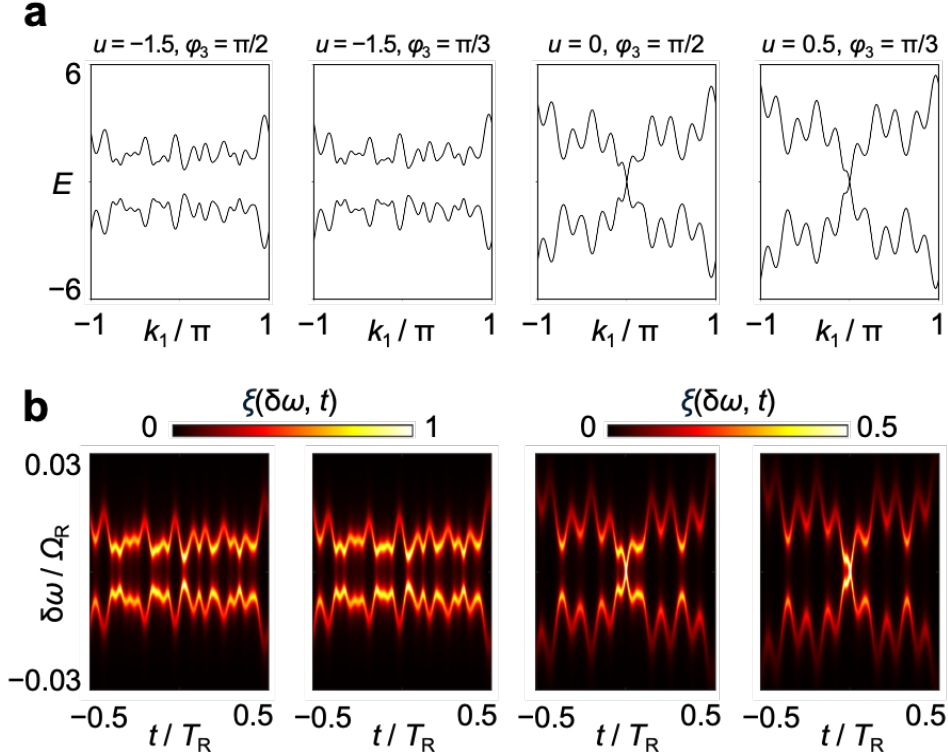

Fig. 3. Simulations of band structure measurements of $H_{3D}$. (a) Theoretical band energies along the one-dimensional sampling line $k_2 = 3k_1 + \varphi_2$ (mod $2\pi$), $k_3 = 9k_1 + \varphi_3$ (mod $2\pi$) in the Brillouin zone by diagonalizing $H_{3D}$. (b) Simulated results of the time-dependent spectrum $\xi(\delta\omega, t)$ of the resonator. We take $\kappa T_R = 0.1$, and the roundtrip power loss of the resonator as 0.07 dB. The spectra $\xi$s with input modes $[1, 0]^T$ and $[0, 1]^T$ are simulated separately, and then averaged and normalized to $[0, 1]$. The use of both input modes improves the uniformity of the excitation levels of the bands [61]; see Supplemental Materials for more information. In the third and fourth panels of (b), the color scale is set to $[0, 0.5]$ for better visibility. In both (a) and (b), we take $\varphi_2 = 0$.

To implement $H_{5D}$ we choose $D = 5$, $S = 4$, $G = 6$. Since $H_{5D}$ uses only $\{\gamma_1, \gamma_2, \gamma_3, \gamma_4, \gamma_5, \gamma_6\}$, a subset of all generators of the SU(4) group, only six branches are needed in the implementation. The mode conversions and mode-selective modulations are summarized in Table 2. Fig. 4 shows the band structures of the Hamiltonian $H_{5D}$. In Fig. 4(a), we plot the theoretical band structures along the one-dimensional sampling line: $k_d = 3^{d-1}k_1 + \varphi_d \pmod{2\pi}$, $d = 2, 3, 4, 5$. We keep $\varphi_2 = \varphi_3 = \varphi_4 = 0$ and $u = 0$ in Fig. 4. In the first two panels $b = 0$ and the system exhibits Yang monopoles. When $\varphi_5 = 0$, the one-dimensional sampling line avoids the Yang monopoles, and the band structure is gapped. When $\varphi_5 = \pi/2$, the sampling line touches a Yang monopole, corresponding to the four-fold degeneracy at $k_1 = 0$ in the second panel. Note that both the upper and lower bands are two-fold degenerate when $b = 0$. In the last two panels $b = 0.5$ and the system exhibits Weyl surfaces. When $\varphi_5 = \pi/3$, the sampling line touches the Weyl surface 3, corresponding to the band crossing at $k_1 = 0$, $E = 0$ in the third panel. When $\varphi_5 = \pi/2$, the sampling line touches the Weyl surface 2, corresponding to the band crossings at $k_1 = 0$, $E = \pm 0.5$ in the last panel. In Fig. 4(b), we show the simulation results of $\xi(\delta\omega, t)$ using the mode conversions and modulations in Table 2, and the resonant features in $\xi(\delta\omega, t)$ closely resemble the theoretical band structures in Fig. 4(a).

Table 2. Modulations and mode conversions to implement $H_{5D}$.

| | $\tau_{g,s=1,2}(t)$ | $\tau_{g,s=3,4}(t)$ | $U_g$ |
|---|---|---|---|
| $g = 1$ | $\exp[+i\kappa T_R \sin(\Omega_R t)]$ | $\exp[-i\kappa T_R \sin(\Omega_R t)]$ | $\frac{1}{\sqrt{2}}\begin{bmatrix} 1 & 0 & 0 & -1 \\ 1 & 0 & 0 & 1 \\ 0 & -1 & -1 & 0 \\ 0 & 1 & -1 & 0 \end{bmatrix}$ |
| $g = 2$ | $\exp[+i\kappa T_R \sin(M\Omega_R t + \varphi_2)]$ | $\exp[-i\kappa T_R \sin(M\Omega_R t + \varphi_2)]$ | $\frac{1}{\sqrt{2}}\begin{bmatrix} -i & 0 & 0 & -i \\ 1 & 0 & 0 & -1 \\ 0 & i & i & 0 \\ 0 & 1 & -1 & 0 \end{bmatrix}$ |
| $g = 3$ | $\exp[+i\kappa T_R \sin(M^2\Omega_R t + \varphi_3)]$ | $\exp[-i\kappa T_R \sin(M^2\Omega_R t + \varphi_3)]$ | $\begin{bmatrix} 0 & 1 & 0 & 0 \\ 0 & 0 & 0 & 1 \\ 0 & 0 & 1 & 0 \\ 1 & 0 & 0 & 0 \end{bmatrix}$ |
| $g = 4$ | $\exp[+i\kappa T_R \sin(M^3\Omega_R t + \varphi_4)]$ | $\exp[-i\kappa T_R \sin(M^3\Omega_R t + \varphi_4)]$ | $\frac{1}{\sqrt{2}}\begin{bmatrix} -1 & 0 & 0 & -1 \\ 0 & 1 & 1 & 0 \\ -1 & 0 & 0 & 1 \\ 0 & 1 & -1 & 0 \end{bmatrix}$ |
| $g = 5$ | $\exp[+i\kappa T_R (4 + u - \cos(\Omega_R t) - \cos(M\Omega_R t + \varphi_2) - \cos(M^2\Omega_R t + \varphi_3) - \cos(M^3\Omega_R t + \varphi_4) - \cos(M^4\Omega_R t + \varphi_5))]$ | $\exp[-i\kappa T_R (4 + u - \cos(\Omega_R t) - \cos(M\Omega_R t + \varphi_2) - \cos(M^2\Omega_R t + \varphi_3) - \cos(M^3\Omega_R t + \varphi_4) - \cos(M^4\Omega_R t + \varphi_5))]$ | $\frac{1}{\sqrt{2}}\begin{bmatrix} -1 & 0 & 0 & -1 \\ 0 & -i & -i & 0 \\ -i & 0 & 0 & i \\ 0 & 1 & -1 & 0 \end{bmatrix}$ |
| $g = 6$ | $\exp(+i\kappa T_R b)$ | $\exp(-i\kappa T_R b)$ | $\frac{1}{\sqrt{2}}\begin{bmatrix} 1 & 0 & 0 & -1 \\ 0 & -i & -i & 0 \\ -i & 0 & 0 & -i \\ 0 & 1 & -1 & 0 \end{bmatrix}$ |

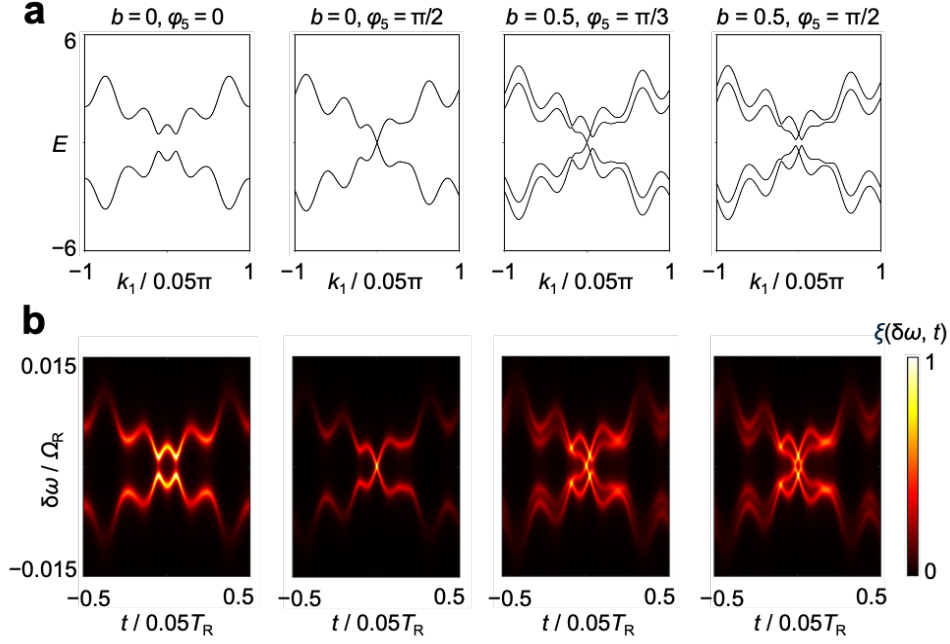

Fig. 4. Simulations of band structure measurements of $H_{5D}$. (a) Theoretical band energies by diagonalizing $H_{5D}$ along the one-dimensional sampling line $k_d = 3^{d-1}k_1 + \varphi_d \pmod{2\pi}$, $d = 2, 3, 4, 5$, in the Brillouin zone. (b) Simulated results of the time-dependent spectrum $\xi(\delta\omega, t)$ of the resonator. We take $\kappa T_R = 0.1$, and the roundtrip power loss of the resonator as 0.03 dB. The spectra $\xi$'s with input modes $[1, 0, 0, 0]^T$, $[0, 1, 0, 0]^T$, $[0, 0, 1, 0]^T$, and $[0, 0, 0, 1]^T$ are simulated separately, and then averaged and normalized to $[0, 1]$. The horizontal axes are zoomed in to $[-0.05\pi, 0.05\pi]$ in (a) and to $[-0.025T_R, 0.025T_R]$ in (b) for better visibility. In both (a) and (b), we take $\varphi_2 = \varphi_3 = \varphi_4 = 0$ and $u = 0$.

**Conclusions**

In this paper, we presented a scheme to explore high-dimensional topological physics using synthetic frequency dimensions in a single ring resonator formed by a multi-mode waveguide. Here the word "multi-mode" can refer to any degree of freedom of the photon other than frequency, including polarization, transverse mode, or orbital angular momentum. We implemented a three-dimensional, two-band model exhibiting Weyl points and topological insulator phases, and a five-dimensional, four-band model exhibiting Yang monopoles and Weyl surfaces as examples. We also showed that the band structures of the models can be measured through the time-dependent spectra of the ring resonator. In this scheme, there is no fundamental limit on the dimensionality or the number of bands of the implementable Hamiltonians. Various Hamiltonians can be realized using the same platform by programming the mode-selective modulations and mode conversions. We expect that this scheme is also applicable to multi-dimensional, multi-band lattice models with non-Hermitian topologies [42,60,64,65].

**Acknowledgements**

The research is supported by MURI projects from the U.S. Air Force Office of Scientific Research (Grants No. FA9550-22-1-0339). C. R.-C. is supported by a Stanford Science Fellowship.

**Declaration of Interests**

The authors declare no competing interests.